\documentclass[aps,prl,showpacs,showkeys,twocolumn,superscriptaddress,
	      floats,floatfix,preprintnumbers]{revtex4-1}
\usepackage{amssymb,amsmath}
\usepackage{bm,url,graphics,subfigure}
\usepackage{epsfig}
\usepackage{ulem}
\usepackage{color}

\def \Jbar {{\overline {\bm J}}}
\def \Vbar {{\overline {\bm V}}}

\def \Nbar {{\overline N}}

\def \nzero {\Nbar_0}
\def \Jbarz {{\overline {J_z}}}
\def \Vbarz {{\overline {V_z}}}

\def \ff {\bm{f}}
\def \xx {\bm{x}}
\def \XX {\bm{X}}
\def \VV {\bm{V}}
\def \vv {\bm{v}}
\def \uu {\bm{u}}
\def \Xdot {\dot{\XX}}
\def \Vdot {\dot{\VV}}

\def \N
\def \Np {N_{\rm p}}

\def \grad {{\bm \nabla}}

\def \delt {\partial_t}
\def \Dt {D_t}

\def \p  {\phi}

\newcommand{\ddz}[1]{\frac{d #1}{dz}}

\newcommand{\thav}[1]{\langle #1\rangle_{{\rm th}}}

{}

\def \mSo {\sigma_{\rm t}}

\def \Rey  {\mbox{Re}}
\def \St {\mbox{St}}
\def \Tu {\mbox{Tu}}
\def \D {\mathcal{D}}
\def \DT {\mathcal{D}_{\rm T}}
\def \n {N}
\def \p {p}

\def \Lx {L_{\rm x}}
\def \Ly {L_{\rm y}}
\def \Lz {L_{\rm z}}

\def \K {\mathcal{K}}
\def \kf  {k_{\rm f}}
\def \qF {q_{\rm F}}

\def \urms  {u_{\rm rms}}

\def \cs  {c_{\rm s}}
\def \kf  {k_{\rm f}}

\def \taup {\tau}

\def \uz {u_{z}}

\def \Mn {\mathcal{N}}
\def \kappat {\kappa_{\rm t}}

\def \tN  {\tt{N}}
\def \tNp  {\tt{N}_{\tt{p}}}
\def \MJ {\mathcal{J}}
 
\def \tauf {\tau_{\rm f}}

\newcommand{\eq}[1]{(\ref{#1})}

\newcommand{\Eq}[1]{Equation~(\ref{#1})}

\newcommand{\fig}[1]{fig~(\ref{#1})}

\def \pre {Phys. Rev. E}
\def \apj {Astrophys. J.}

\begin{document}
\title{Turbophoresis in forced inhomogeneous turbulence}
\author{Dhrubaditya Mitra}
\affiliation{Nordita, KTH Royal Institute of Technology and Stockholm University,
Roslagstullsbacken 23, 10691 Stockholm, Sweden}
\author{Nils Erland L. Haugen}
\affiliation{Department of Energy and Process Engineering, NTNU, 7491 Trondheim, Norway}
\affiliation{SINTEF Energy Research, 7465 Trondheim, Norway}
\author{Igor Rogachevskii}
\affiliation{Department of Mechanical Engineering, Ben-Gurion University of the Negev,
POB 653, Beer-Sheva 84105, Israel}
\affiliation{Nordita, KTH Royal Institute of Technology and Stockholm University,
Roslagstullsbacken 23, 10691 Stockholm, Sweden}
\date{\today,~  }
\begin{abstract}
We show, by direct numerical simulations,  that
heavy inertial particles (characterized by Stokes number $\St$)
in inhomogeneously forced statistically stationary
isothermal turbulent flows cluster at the minima of
mean-square turbulent velocity.
Two turbulent transport processes,
turbophoresis and turbulent diffusion together determine
the spatial distribution of the particles.
If the turbulent diffusivity is assumed to scale
with turbulent root mean square velocity,
as is the case for homogeneous turbulence,
the turbophoretic coefficient can be calculated.
Indeed, for the above assumption, the non-dimensional product of the
turbophoretic coefficient and the rms velocity is shown to
increase with $\St$ for small $\St$, reach a maxima
for $\St\approx 10$ and decrease as $\sim \St^{-0.33}$
for large $\St$.
\end{abstract}
\preprint{NORDITA-2016-15}
\maketitle
\section{Introduction}
The subject of this paper is turbulent flows of Newtonian fluids with
small particles.  We limit ourselves to spherical particles whose
radii are significantly smaller than the viscous scale of the
fluid. We further assume that the material density of each particle is
significantly larger than that of the fluid and the number density of
the particles is quite low.
In this parameter regime the equations of
motion of each individual particle are well described by
\begin{eqnarray}
\Xdot &=& \VV  \nonumber\\
\Vdot &=& \frac{1}{\tau}\left\{ \uu[\XX(t),t] - \VV \right\} \/,
\label{eq:HIP}
\end{eqnarray}
where $\XX$ and $\VV$ are respectively the position and velocity
of the particle,
$\tau$ is the characteristic relaxation time of the particle
and $\uu$ is the flow velocity,
whose magnitude is denoted by $u$.
As the number density of the particles is assumed to be small,
the effects of the particles on the flow and particle-particle
interactions are both ignored.
The flow velocity $\uu$ is obtained by solving the Navier--Stokes
equation with proper boundary conditions. Particles that satisfy
\eq{eq:HIP} are called Heavy Inertial Particles (HIP).
This approximation is often used to model various natural
and industrial flows. Two important examples are:
(a) droplets in warm clouds \cite[see, e.g.,][for a review]{sha03,gra+wan13} and
(b) dust grains in astrophysical accretion
disks \cite[see, e.g.,][for a review]{Arm10}.

The study of heavy inertial particles in homogeneous and isotropic
turbulent (HIT) flows is a topic of
numerous research publications.
In addition to the fluid Reynolds number,
such a problem is described by
a second dimensionless number, the Stokes number, defined by
$\St \equiv \tau/\tauf$,
where $\tauf$ is a characteristic time scale of the flow.
Recent analytical, numerical, and experimental works have shown that
HIPs self-organize in fractal clusters in HIT, \cite[see, e.g.,][and
references therein]{Zai+ali+sin08,tos+bod09} and at small distances
the probability distribution function (PDF) of their relative
velocities shows power-law behavior~\cite{gus+meh11,per+jon15}.

The focus of this paper is the large-scale clustering of heavy inertial particles
in {\it inhomogeneous}, but unstratified, turbulence.
Numerical and experimental
studies in wall-bounded flows, e.g., channel and pipe flows,
have shown that heavy inertial particles distribute themselves
preferentially near the wall -- a phenomenon that has been called
turbophoresis~\cite{cap+tam+tro+vit75}
by analogy with thermophoresis~\cite{Gro+maz62,duh+bra06_prl,duh+bra06_pnas}.
Thermophoresis -- motion of Brownian particles governed by gradient in
temperature~\cite{max1879} -- is well understood within the framework of
local thermodynamic equilibrium in  statistical mechanics~\cite{Gro+maz62}.
Recent works~\cite{duh+bra06_prl,duh+bra06_pnas} have demonstrated that the
flux of molecules in thermophoresis ($\MJ$) can be well described by
\begin{equation}
\thav{\MJ} = -\D \grad \thav{\Mn} -\DT \thav{\Mn} \grad T
\label{eq:tph}
\end{equation}
where $\D$ is the diffusion coefficient, $\Mn$ the concentration of the molecules,
$\DT$ the thermophoretic coefficient, and
the symbol $\thav{\cdot}$ denotes averaging over a
state in local thermal equilibrium at temperature $T$.

To understand turbophoresis, let us consider a statistically
stationary inhomogeneous turbulent flow. For simplicity, we consider
the turbulence to be inhomogeneous in only one coordinate direction,
which we denote by $z$.
This is often the case in most practical applications, e.g., in turbulent
channel or pipe flows. We start with the mean flux ($\Jbar$) of inertial
particles in space,
\begin{equation}
\Jbar \equiv \overline {\n \VV} \/,
\label{eq:flux1}
\end{equation}
where the symbol $\overline{(\bullet)}$ denotes Reynolds averaging,
and $\n$ denotes the number density of the particles.
We use the standard Reynolds decomposition,
\begin{equation}
\n = \Nbar + n \/,\hspace{1cm} \VV = \Vbar + \vv  \/,
\end{equation}
where each quantity of interest is decomposed into
mean and fluctuating parts.
After Reynolds averaging we obtain
\begin{equation}
\Jbar = \Nbar\hspace{0.1cm}\Vbar + \overline{n \vv} \/.
\label{eq:flux2}
\end{equation}
The second term on the right hand side (RHS) of this equation
is usually modelled through closure by
$\overline{n \vv} = -\kappa \nabla \Nbar$, where $\kappa$ is the
turbulent diffusivity. To calculate the first term in the RHS of
\eq{eq:flux2} we use \eq{eq:HIP} to write (following
Ref.~\cite{max87_approx})
\begin{eqnarray}
\VV &=& \uu - \taup \Vdot \nonumber \\
       &\approx& \uu - \taup \biggl[{\partial \uu
\over \partial t} + (\uu \cdot \grad) \uu
\biggr] + {\rm O}(\taup^2)\/,
\end{eqnarray}
which is an approximation valid for small $\taup$.
On Reynolds averaging, we obtain
\begin{equation}
\Vbar \approx \overline{\uu} - \taup {\partial \overline{\uu}
\over \partial t} - \taup \overline{(\uu \cdot \grad) \uu}.
\label{eq:Vbar}
\end{equation}
In general, we are interested in the component of the flux
$\Jbar$ along the direction of inhomogeneity,
which in wall bounded flows, e.g., pipe or channel flows,
is the wall-normal direction. Projecting \eq{eq:Vbar}
along the wall-normal direction we obtain
\begin{equation}
\Vbarz \approx - \taup \ddz{ \overline{(\uz^2/2)}}\/,
\end{equation}
because $\overline{u_z} = 0$.
This demonstrates that although the mean flow velocity in the
wall normal direction is zero, the mean particle velocity is not.
Putting together, the flux of inertial particles along the direction
of inhomogeneity is given by
\begin{equation}
\Jbarz \approx -\taup \Nbar \ddz {\overline{(\uz^2/2)}} - \kappa \ddz{ \Nbar}.
\label{eq:flux3}
\end{equation}
The second term on the RHS is the usual Fickian (turbulent) diffusion, the first
one is called the turbophoretic term.
There are several ways to obtain \eq{eq:flux3}, here we have followed
Refs.~\cite{guha97,elp+kle+rog98,guha08}.
Clearly, \eq{eq:flux3} is an
approximation that holds for small $\St$ (small $\taup$).
How can this expression be generalized
for arbitrary values of the Stokes number ?

In this paper we show by direct numerical simulations (DNS)
that the large-scale clustering of heavy inertial particles in
inhomogeneous turbulence can be described by a
simple generalization of \eq{eq:flux3};
\begin{equation}
\Jbar =  - \kappat \Nbar \grad \K -\kappa \grad \Nbar \/.
\label{eq:tuph}
\end{equation}
Here  $\kappat$ is the {\it turbophoretic} coefficient,
$\kappa$ is turbulent diffusivity,
and $\K \equiv \overline{(u^2/2)}$.
The expression of molecular flux in thermophoresis, \eq{eq:tph},
can be considered as an inspiration for \eq{eq:tuph}.
So far, neither numerical nor experimental studies have
measured $\kappat$ and $\kappa$ simultaneously
and it is not clear how this can be done.
Here, we perform careful numerical experiments
to measure their ratio:
\begin{equation}
\Tu \equiv \frac{\kappat}{\kappa},
\label{eq:ratio}
\end{equation}
which we call the turbophoretic ratio.
We find that our data is
well described by $\Tu = \mSo/(2\sqrt{\K})$.
We non-dimensionalize $\mSo$ by multiplying it with $\urms$;
the non-dimensionalized $\mSo$
is a non-monotonic function of the Stokes number -- it increases
linearly with  $\St$  for small
$\St$, reaches a maximum value around $\St \approx 10$ and decreases with $\St$ for
large $\St$~\footnote{Note that we define the Stokes number using
the characteristic time scale of the flow, $\tauf$,
based on the integral (forcing) scale of turbulence.
Very often the Stokes number is defined using
the flow time scale based on the Kolmogorov scale.}.
This is the first numerical measurement of the turbophoretic ratio
and its dependence on the Stokes number.

\section{Model}
So far, most numerical~\cite{mcl89,bro+kon+han+mcl92,mar+sol02,sar+sch+bra+pic12}  and
experimental~\cite{liu+aga74} observations
of large-scale clustering  and deposition~\cite{seh70,seh71} of HIP in
inhomogeneous flows are in wall-bounded
flows~\cite[see, e.g.,][for a review]{guha08}.
In such flows it has not been possible to
disentangle the individual effects of inhomogeneities of mean-square turbulent velocity,
large-scale shear, and physical boundaries.
For a deeper understanding of turbophoresis, we choose to do DNS of
the Navier--Stokes equation
\begin{eqnarray}
\Dt \rho &+& \rho \partial_k u_k = 0 ,\nonumber \\
\rho \Dt u_j&=& -\partial_j \p + \mu \partial_k S_{kj} + f_j F(z) ,
\label{eq:fluid}
\end{eqnarray}
under isothermal conditions, with an inhomogeneous
external force.
Here $\partial_k$ denotes partial derivatives with respect to
coordinate $k$, $\Dt \equiv \delt + u_k\partial_k$ is the advective
derivative,  $\uu$ (whose $k$-th component is $u_k$), $\p$, and $\rho$ are
respectively the velocity, pressure, and density of the flow,
$\mu$ is the dynamic viscosity, and
$S_{kj} \equiv \partial_k u_j + \partial_j u_k -\delta_{jk}(2/3)\partial_k u_k$
is the viscous stress.
The simulations are performed in a three-dimensional periodic box
with sides $\Lx=\Ly=\Lz=2\pi$.
In addition we use the ideal gas equation of state with a constant
speed of sound $\cs = 1$.

The flow attains a statistically stationary state where
the average energy dissipation by viscous forces is balanced by
the average energy injection by the external force $\ff$ which is random,
Gaussian, white-in-time, concentrated on a shell of wavenumber with
radius $\kf$ in Fourier space~\cite{B01}.
The amplitude of the external force is chosen such that the Mach number
is always less than $0.1$, i.e., the flow is weakly compressible.
Crucially, the flow is made inhomogeneous (but statistically isotropic) by the function
$F(z)$ which is a function of one coordinate
direction, $z$.  We choose the function $F(z)$ to be a slowly varying
function of $z$, i.e., the characteristic wavenumber of $F(z)$, $\qF \ll \kf$.
In particular, we examine three cases:
(A) $F(z) = \exp[-(z/H)^2]$ with $H=1$, (B)  $F(z) = \sin^2(z)$, and (C) $F(z) = \sin^2(2z)$.
In the context of the present problem, length scales large than $1/\kf$ are
the large scales.

We use the pencil-code~\cite{pencil-code},
which uses a sixth-order finite-difference scheme for space
derivatives and a third-order Williamson-Runge-Kutta~\cite{wil80} scheme for time
derivatives.  The same setup, with a homogeneous external force, has been used in
studies of scaling and intermittency  in fluid and magnetohydrodynamic
turbulence~\cite{dob+hau+you+bra03,hau+bra+dob03,hau+bra04}.
We introduce the particles into the DNS
after the flow has reached statistically stationary state.
Then we simultaneously solve the equations of the flow, \eq{eq:fluid}, and the
HIPs, \eq{eq:HIP}.
To solve for the HIPs in the flow we have to interpolate the flow velocity to
typically off-grid positions of the HIPs. We use a tri-linear method for
interpolation.

We define the Reynolds number by $\Rey\equiv \urms/(\nu\kf)$,
where $\urms$ is the root-mean-square velocity of the flow
averaged over the whole domain and the kinematic
viscosity $\nu =\mu/\rho$.
As we are interested in large-scale clustering, we use
the large-eddy-turnover time,
$\tauf \equiv 1/(\urms\kf)$, which is the
characteristic time scale of the flow at scale $1/\kf$,
to define the Stokes number. In our simulations we
use three different value of $\Lx\kf/2\pi=5,15$, and $30$.
Here we implement Reynolds averaging by averaging over the $x$ and $y$
coordinate directions.
On top of the Reynolds averaging, we also perform time averaging
over several thousand large-eddy turnover times.
The various relevant parameters of our simulations are
shown in Table~\ref{table:para}.

\section{Results}
\begin{figure}[h]
\begin{center}
\includegraphics[width=0.90\columnwidth]{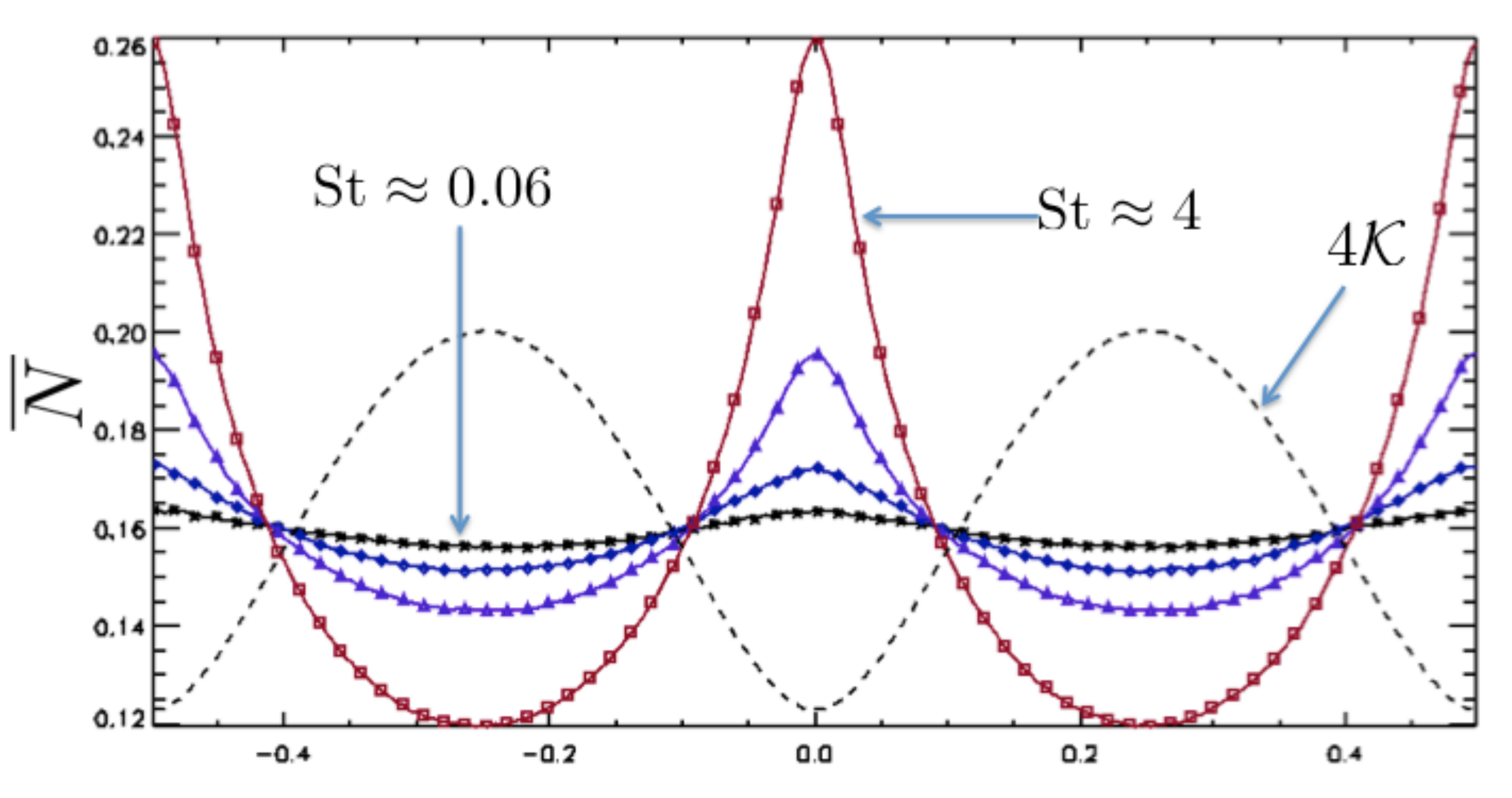}
\includegraphics[width=0.90\columnwidth]{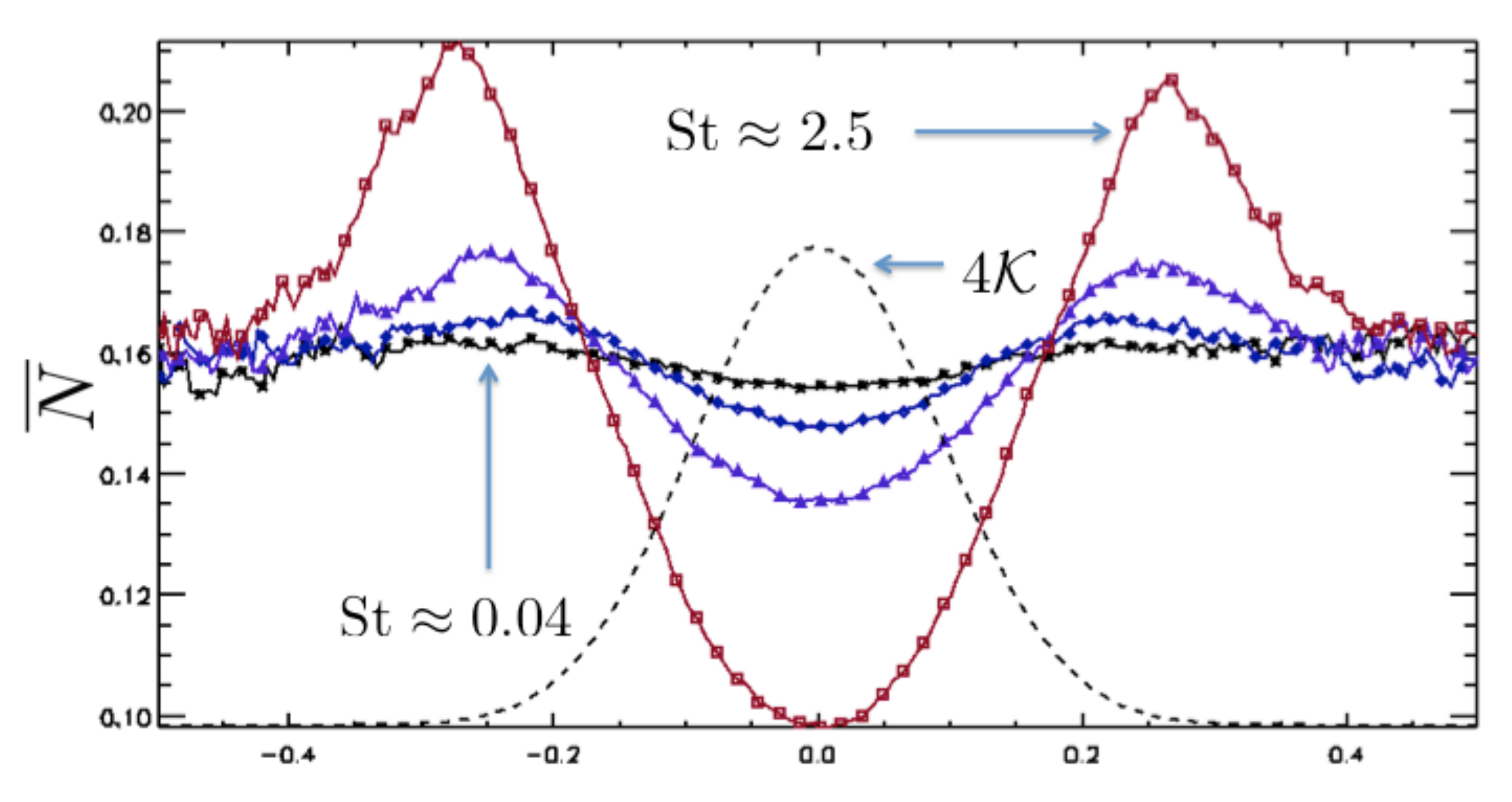}
\includegraphics[width=0.90\columnwidth]{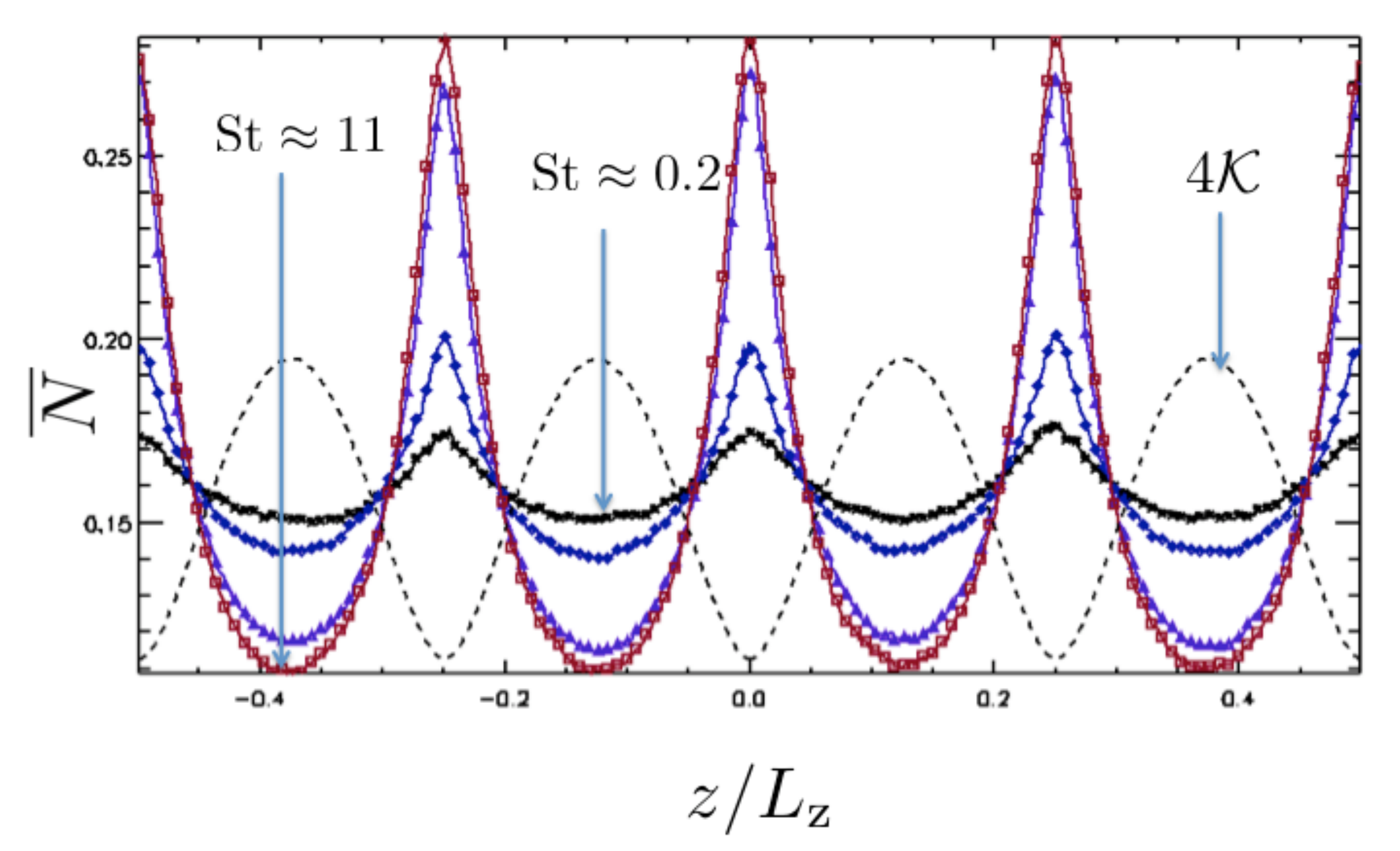}
\caption{\label{fig:np}
The number density of the particles averaged over the $x$ and $y$
coordinates, $n(z)$, as a function of $z/\Lz$ ($\Lz=2\pi$) for
(top) $F(z) = \sin^2(z)$ and $\St \approx 0.06, 0.25, 1.0,$ and
$4.0$, (middle) $F(z) = \exp(-(z/H)^2)$, and $\St \approx 0.04, 0.16, 0.64$, and
$2.6$, (bottom) $F(z) = \sin^2(2z)$ and $\St \approx 0.17, 0.7, 2.8$, and
$11$. The dashed lines shows the mean-squre turbulent velocity, $\K(z)$,
multiplied by a factor of four to fit it in the same scale.
}\end{center}
\end{figure}

Once our simulations reach a statistically stationary state
we plot, in \fig{fig:np}, the mean-square turbulent velocity
$\K(z) \equiv (1/2)\overline{u^2}$ and
the mean number density of the particles, $\Nbar$,
for the three different $F(z)$ and
for different values of the Stokes number.
Here the Eulerian number density of the particles
$\n(\xx) \equiv \delta^3(\xx-\XX)$,
where $\delta^3(\cdot)$ is the three-dimensional Dirac delta function.
Clearly, the particles cluster away from the maxima of $\K$, i.e.,
turbophoresis is observed.
\begin{figure}[h]
\begin{center}
\includegraphics[width=0.90\columnwidth]{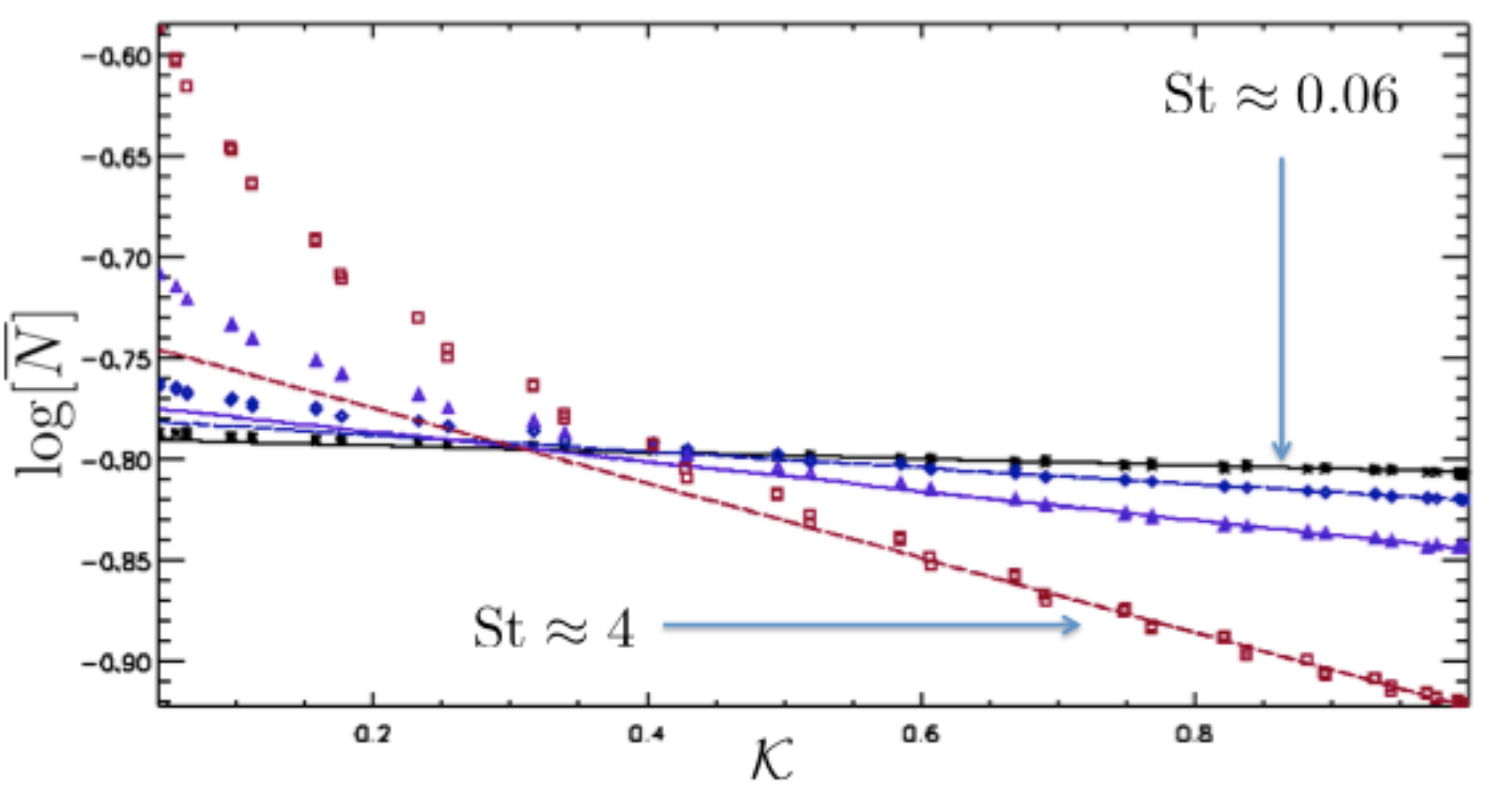}
\includegraphics[width=0.90\columnwidth]{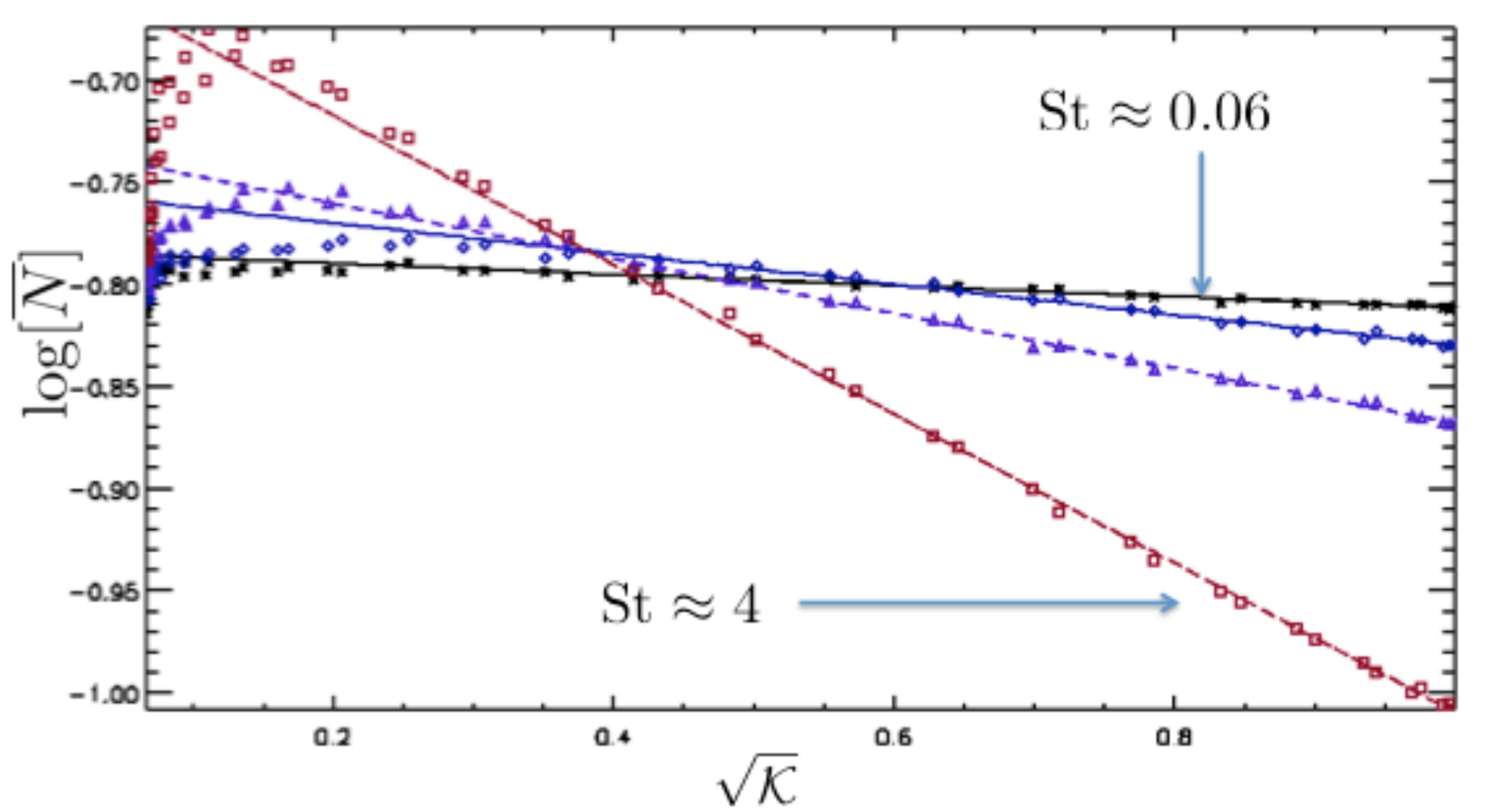}
\includegraphics[width=0.90\columnwidth]{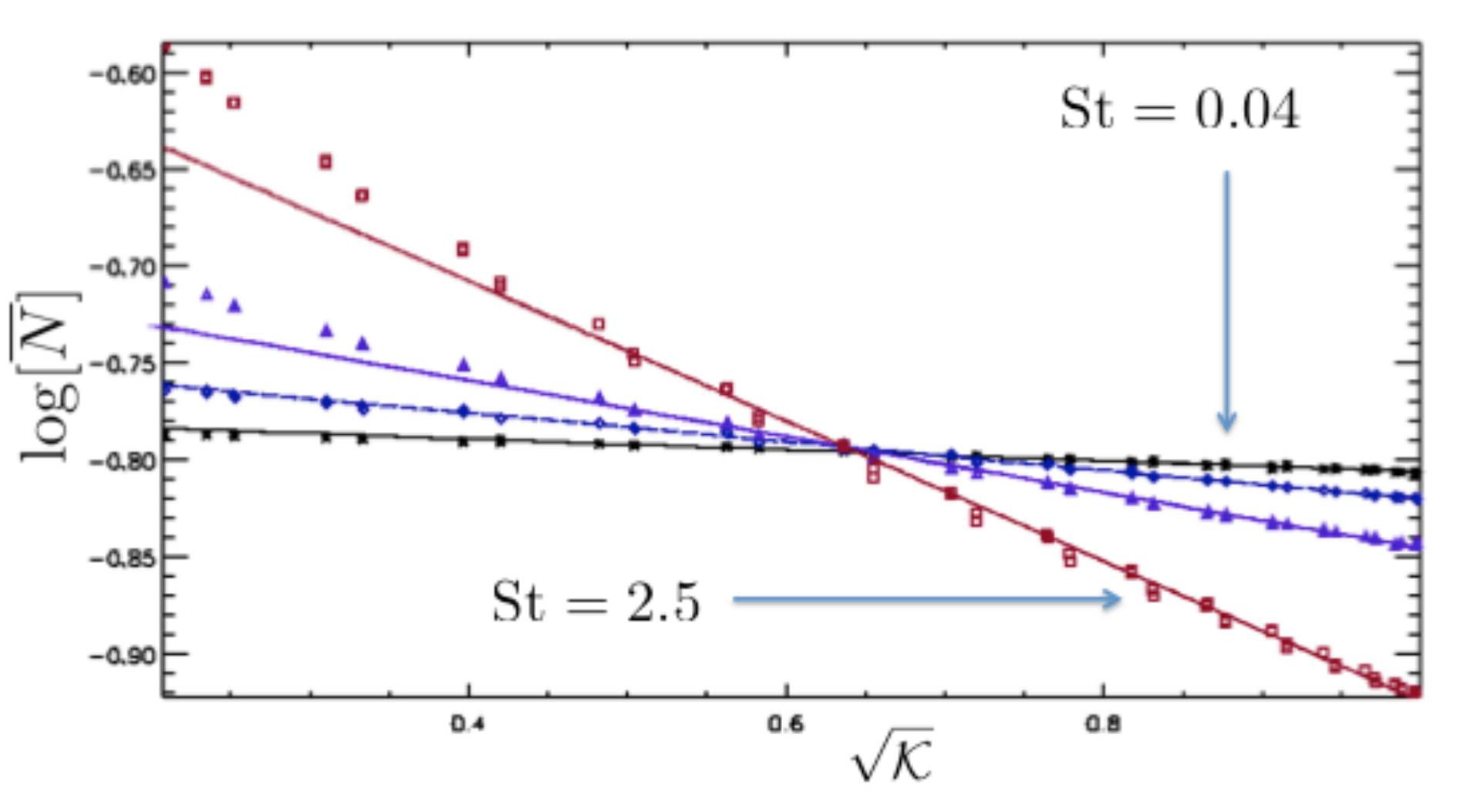}
\caption{\label{fig:fit}
(top) Plot of $\log(n)$ as a function of the mean-square turbulent velocity $\K$
for $F(z) = \sin^2(z)$ and $\St \approx 0.06, 0.25, 1.0,$ and
$4.0$ which corresponds to the data shown in the top panel of \fig{fig:np}.
The data is fitted with a straight line following \eq{eq:con1}.
(middle) The same data now plotted as a function of $\sqrt{K}$ and fitted
with \eq{eq:con2}. Compare this plot with the top panel.
(bottom) The same plot for $F(z) = \exp(-(z/H)^2)$,
and $\St \approx 0.04, 0.16, 0.64,$ and
$2.5$, which corresponds to the middle panel of \fig{fig:np}.
}\end{center}
\end{figure}

We now want to quantify turbophoresis using \eq{eq:tuph}.
At stationary state, the mean flux, $\Jbar$,  must be
zero~\footnote{Here as in elsewhere in the paper we ignore the molecular diffusivity
of the particles,
i.e., the ratio of the kinematic viscosity to the diffusion coefficient
(the Schmidt number) is very large. This is a typical condition for HIP}.
Furthermore, as the turbulence is inhomogeneous in only one direction, $z$,
\eq{eq:tuph} reduces to the following one dimensional problem:
\begin{equation}
\ddz{\Nbar} + \frac{\kappat}{\kappa}\Nbar \ddz{\K} = 0.
\label{eq:1d}
\end{equation}
Clearly, it is impossible to measure the turbulent transport
coefficients $\kappat$ and $\kappa$ individually.
Instead, only their ratio,
which we call the turbophoretic ratio, $\Tu$,
can be measured.
In addition, to integrate \eq{eq:1d} and compare with our data we need
to make conjectures about the spatial dependence of $\Tu$.
The naive one would be to assume that $\Tu$ does not depend
on space; which would imply
\begin{equation}
\Nbar = \nzero \exp(-\Tu \K).
\label{eq:con1}
\end{equation}
In the top panel of \fig{fig:fit} we show that this conjecture does not
fit our data well even in the best of cases.
Let us note now, that although
we do not know exactly what $\kappa$ would
be in inhomogeneous turbulence, the turbulent diffusion has been a well
studied problem in homogeneous and isotropic turbulence,
in which case $\K$ is not a function of space and one typically obtains
$\kappa \sim \sqrt{K}$.
In the present problem $\K$ is a slowly varying function of space, hence
we conjecture that $\kappa(z) \sim \sqrt{\K(z)}$ is also a slowly
varying  function of space, i.e.,
$\Tu \equiv  \mSo/(2\sqrt{\K})$ with $\mSo$ a constant in space.
With this substitution, \eq{eq:1d} can be integrated
to obtain
\begin{equation}
\Nbar = \nzero \exp(-\mSo \sqrt{\K})\/,
\label{eq:con2}
\end{equation}
which fits our numerical data quite well as we show
for two different cases in the middle and the bottom panel of  \fig{fig:fit}.
If $\kappa \sim  \sqrt{\K}$ were true even in inhomogeneous
turbulence we would have obtained $\mSo \sim \kappat$, i.e., the coefficient
$\mSo$ is proportional to the turbophoretic coefficient $\kappat$.

\begin{figure}[h]
\begin{center}
\includegraphics[width=0.90\columnwidth]{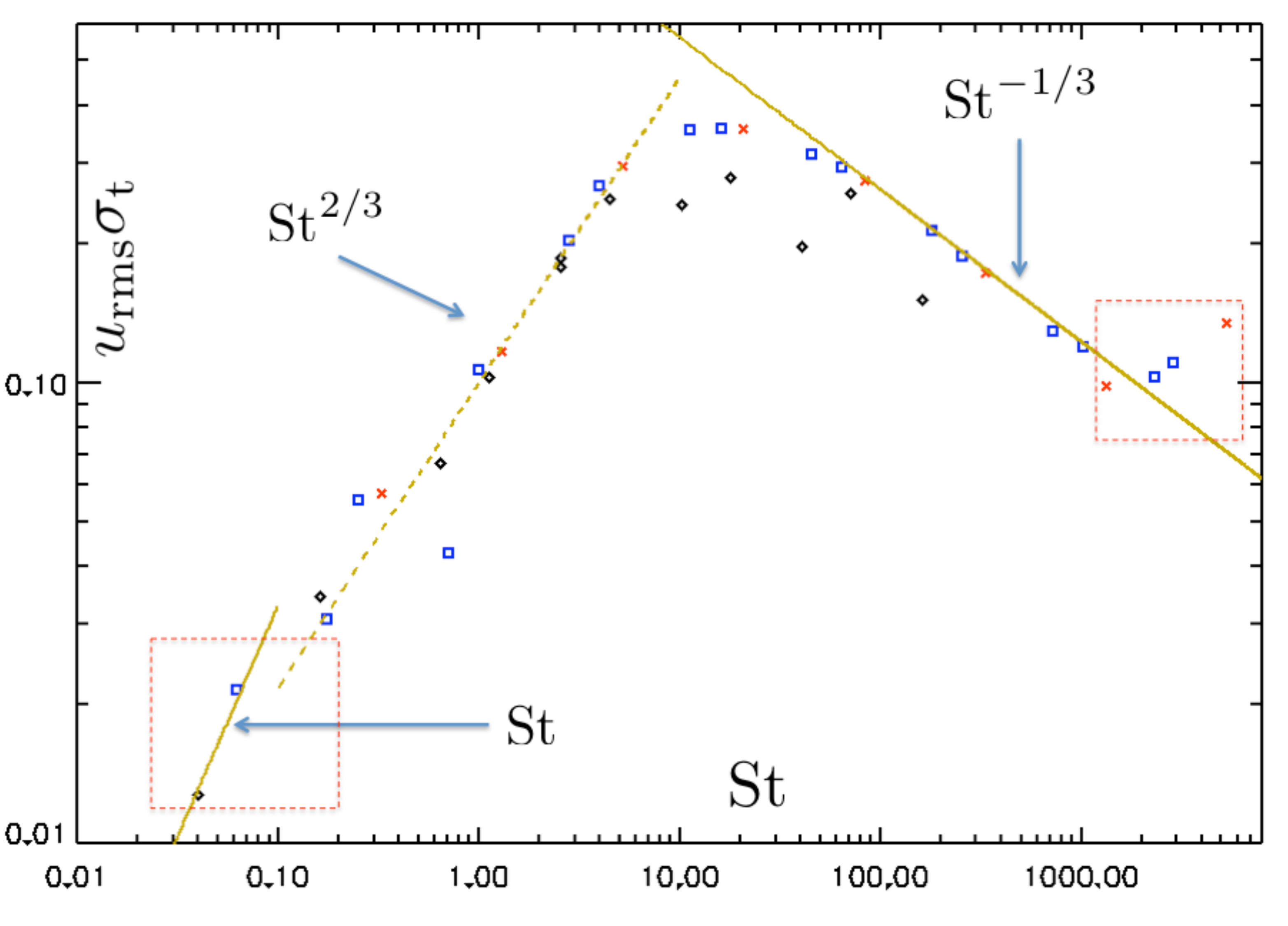}
\caption{\label{fig:SoSt}
The dependence of the non-dimensional quantity
$\urms\mSo$ on
Stokes number from all of our runs. The symbols
$\diamond$, $\Box$, $\times$, denote the runs
$A$ $B$ and $C$ in table~\ref{table:para} respectively.
Enclosed within a dashed box are those measurements with large error.
}\end{center}
\end{figure}

We calculate $\mSo$ for a range of values of $\St$ for three different
spatial distributions of the mean-square turbulent velocity,
 three different values of $\Lx\kf/2\pi$, two different
resolutions ($256^3$ and $512^3$) and for several values of
the Reynolds number ranging from $12$ to $56$.
The results, non-dimensionalized by calculating
$\mSo\urms$,
are plotted in \fig{fig:SoSt}. For very small and
very large values of $\St$ our results are not very precise because
the large-scale clustering is weak for both of these limits. These data points are
marked in \fig{fig:SoSt} with a dashed box around them.
Nevertheless, for very small $\St$ we find that $\mSo$ is linearly
proportional to $\St$.
For moderate values of $\St$ we find that $\mSo$ increases
with $\St$ roughly as $\St^{2/3}$.
At around $\St\approx 10$ this trend reverses and $\mSo$
{\it decreases} with $\St$, in other words the clustering
phenomenon becomes weaker. For higher values of $\St$
we find that $\mSo$ decreases approximately as $\St^{-1/3}$.

To summarize, two results lie at the heart of our paper:
(A) We quantify large-scale clustering in inhomogeneous
turbulent flows through two turbulent transport coefficients
turbulent diffusion and turbophoresis, whose ratio we call the
turbophoretic ratio, $\Tu$. The turbophoretic ratio, $\Tu$ is a
function of space such that the function $\mSo \equiv (2\sqrt{\K})\Tu$ is
a constant.
(B) We find that the $\mSo$ (non-dimensionalized by multiplying with $\urms$) 
is a non-monotonic function  of Stokes number
with  maximum clustering appearing at $\St \approx 10$.

\section{Discussions}
Large-scale clustering of inertial particles in inhomogeneous
turbulent flows is commonly observed in geophysical and
engineering  problems.
But in trying to understand such
clustering we encounter a difficult problem due to
an inhomogeneous distribution of mean-square turbulent velocity,
presence of large-scale shear, anisotropy, gravity
and  boundary layers.
To make progress, it is essential that we study each effect
in isolation. To this end, in this paper, we have proposed
a model in which large-scale clustering is due solely to the
inhomogeneous distribution of mean-square turbulent velocity.
We find that the minimum of the number density of the particles
appears at the maximum of the mean-square turbulent velocity.
Furthermore, we find that this clustering is determined by two processes:
turbophoresis and
turbulent diffusion. We numerically measure the ratio of these two transport
coefficients, which we call the turbophoretic ratio, $\Tu$,
which is given by
$\Tu = \mSo/(2\sqrt{\K})$ with $\mSo$
is an universal function of $\St$,
increasing with $\St$ for small $\St$ reaching a peak near $\St \approx 10$
and then decreasing with $\St$.

Traditionally such problems are studied by first constructing
exact, but unclosed, relations between various moments of the
probability distribution functions of flow velocity and
density and velocity of the particle phase and then imposing
arbitrary closures to obtain expressions for transport coefficients, see, e.g.,
Refs~\cite{elg94,fox14,cap+des+fox15}.
In this paper we have taken a different approach. We have assumed
that the large scale transport is determined by two undertermined turbulent
transport coefficients and then numerically calculated the ratio of the
transport coefficients as a function of the Stokes number.

A special mention must be made of Ref.~\cite{del+cen+mus+bof15}
where the same clustering phenomenon is studied in turbulent
Kolmogorov flows. Although the authors do not interpret their results
as a balance between turbophoretic and turbulent diffusive fluxes as we do,
they do observe that clustering increases for small $\St$ but this
trend reverses smoothly at higher values of $\St$.

In the introduction we have provided an approximate analytical
theory for turbophoresis which gives a linear dependence of the
turbophoretic coefficient on $\St$ for small $\St$.
To the best of our knowledge, this result was first
obtained in Ref.~\cite{ree83}, using the Lagrangian History
Direct Interaction Approximation, \cite[see also][for a review]{reeksparticle}.
Our data does reproduce this
trend for very small $\St$ although for those values of $\St$
the clustering is quite weak and the errobars on $\mSo$ are large.

An alternative way to analytically understand turbophoresis
would be to replace the fluid velocity, $\uu$, in \Eq{eq:HIP} by
a Gaussian white noise whose strength is the function $F(z)$ of
one coordinate, $z$.
Treating \Eq{eq:HIP} as a Langevin problem it is then possible
to write the corresponding Fokker--Planck equation which is an
equation for the probability distribution function $p(z,v)$.
The particle number density can be obtained by $\n(z) = \int p(z,v) dv$.
For a general $F(z)$ a stationary solution of this Fokker--Planck
equation is not known.
If the strength of the noise $F(z)=c$  is taken to be a constant,
then it is possible to obtain a flux-free stationary solution of the
Fokker--Planck equation to obtain $p(z,v) \sim \exp(-\tau v^2/c^2)$.
If $F(z)$ is assumed to be a slowly varying function of space then
using the method of multiple scales it is possible to calculate a flux
that has both a diffusive and a turbophoretic part~\cite[see the
supplemental material of Ref. ][]{cel+bo+eic+aur12}. But in such
a model the turbophoretic ratio
is no longer a function of $\St$ but a constant.
To capture the non-monotonic dependence of
$\Tu$ on $\St$, as we observe, it may be necessary to use a
colored noise instead of white-in-time noise.
In Ref.~\cite{bel+fou+fal14}  it has been shown that for $F(z) = z^2$
it is possible to find a
non-trivial solution of the Fokker-Planck equation, which leads to a
localization-delocalization transition.
In other words no large-scale clustering should
be observed above a critical value of the Stokes number.
This is not found in our numerical experiments.

The large-scale clustering observed in turbulent pipe flows~\cite{noo+sar+bra+sch15},
which is similar to what has been observed in
turbulent channel flows in Ref.~\cite{sar+sch+bra+pic12},
can be analyzed in a  similar manner to extract
the turbophoretic ratio
whose dependence on the Stokes number is similar
to what has been observed by us but for the data very close to the
wall and near the center of the pipe~\footnote{A. Noorani,
private communication.}.
We believe that to understand such departure the effects of
the boundary layer and anisotropy has to be taken into account.

In our simulations we have assumed that the motion of the particles is well
described by the HIP approximation, \eq{eq:HIP}.
For small $\St$ the validity of this approximation has been critically
examined, by comparing DNS and experimental data for
homogeneous and isotropic turbulence, in Ref.~\cite{saw+bew+bod+ray+bec14}.
To the best of our knowledge no such study exists for large $\St$ and
for inhomogeneous turbulence. It may turn out that the large $\St$
behavior of $\mSo$ shown in \fig{fig:SoSt} may not be
easily realisable in experiments. Nevertheless, the non-monotonic
behavior of $\mSo$ as a function of $\St$ with a peak at around $\St\approx 10$
should be experimentally realisable.
Note further that we have assumed that the particles are
passive, which implies that our results apply to experiments where
the mass loading parameter is small. Clustering of particles is also
found in simulations where the feedback from the particles to
the flow is included, see, e.g., Refs.~\cite{fox14,cap+des+fox15}.
In those cases, gravity may also play an important role, i.e.,
horizontal and vertical pipes may show different clustering
phenomena, see. e.g., Refs.~\cite{vin06,uhl08,lel+big+dop+vin+cha09,cap+des+fox16}.
A description of such clustering through turbophoresis and turbulent
diffusion can also be done but is outside the scope of this paper.

In this work, we have considered an isothermal flow which is relevant
to most engineering applications.
In geophysical problems, it is common to have
fully developed turbulent flows with an inhomogeneous
distribution of mean temperature.
In such cases,  the HIPs are found to accumulate
in the vicinity of the minima of mean temperature.
This large-scale clustering phenomenon, which can be described
by turbulent thermal diffusion, was predicted in analytical study
\cite{EKR96,EKR97,elp+kle+rog98,EKR00,EKR01}
and detected in laboratory experiments
\cite{BEKR04,EEKR06,EKR10},
DNS \cite{HKRB12} and atmospheric
observations \cite{SEKR09}.
\begin{table}
\begin{center}
\begin{tabular}{| c| c| c| c| c| c| }
\hline\hline
Run & $\tN$ & $\tNp$                  & $\Rey$ & $\Lz\kf$  &  $\St$ \\ \hline
A1 &   $256$ & $5\times 10^{5}$       & $48$    & $5$          & $1.1 - 71.6$ \\
A2 &   $256$ & $5\times 10^{5}$       & $13$    & $15$         & $0.4 - 162$ \\
B1 &   $256$ & $ 10^{6}$              & $20$    & $15$         & $0.6 - 102.$ \\
B2 &   $256$ & $ 10^{6}$              & $20$    & $15$         & $5. - 1028.$ \\
B3 &   $512$ & $ 10^{6}$              & $56$    & $15$         & $0.17 - 2896.$ \\
C1 &   $512$ & $ 10^{6}$              & $26$    & $30$         & $0.32 - 333.$ \\
\hline
\hline
\end{tabular}
\caption{\label{table:para}Parameters for our DNS runs:
The runs a grouped by the function $F(z)$ which determines the spatial variation of the
mean-square turbulent velocity $\K$, in particular;
(A) $F(z) = \exp[-(z/H)^2]$ with $H=1$, (B)  $F(z) = \sin^2(z)$, and (C) $F(z) = \sin^2(2z)$.
The number of Eulerian grid points are $\tN^3$, the number of HIPs is given by $\tNp$.
}
\end{center}
\end{table}

\section{Acknowledgment}
IR thanks N. Kleeorin and T. Elperin for stimulating discussions.
DM thanks S. Bo, R. Eichorn, S. Mussachio, A. Noorani, and J. Wettlaufer
for useful discussions. A part of the work was done
while DM was visiting Woods Hole Oceanographic Institute
whose hospitality is gratefully acknowledged.
DM is supported by the grant ”Bottlenecks for particle growth in
turbulent aerosols” from the Knut and Alice Wallenberg
Foundation (Dnr. KAW 2014.0048) and by VR grant 638-2013-9243.
NELH and IR are supported by the Research Council of Norway under the FRINATEK
grant 231444. Some of the computations were performed on resources
provided by the Swedish National Infrastructure for Computing (SNIC)
at PDC and at HPC2N.
\printtables
\bibliography{turb_ref,sunref}


\end{document}